\documentclass[pra,twocolumn,floatfix]{revtex4-1}
\usepackage{amsmath,amsfonts}
\usepackage{graphicx}
\usepackage{color}
\usepackage{enumerate}
\usepackage[hidelinks,colorlinks=true,citecolor=blue,urlcolor=blue,linkcolor=blue]{hyperref}

\begin{document}


\title{Decoupled heat and charge rectification as a many-body effect in quantum wires}

\author{Conor Stevenson}
\author{Bernd Braunecker}
\affiliation{SUPA, School of Physics and Astronomy,
	University of St Andrews, North Haugh, St Andrews KY16 9SS, UK}

\date{\today}


\begin{abstract}
We show that for a quantum wire with a local asymmetric scattering potential the
principal channels for charge and heat rectification decouple and renormalise
differently under electron interactions, with heat rectification generally being
more relevant. The polarisation of the rectification results from quantum interference
and is tuneable through external gating. Furthermore, for spin polarised or helical electrons and sufficiently
strong interactions a regime can be obtained in which heat transport is strongly
rectified but charge rectification is very weak.
\end{abstract}


\maketitle


\section{Introduction}

Electronic technology relies significantly on the progressive miniaturisation of its components.
This has started reaching into the quantum regime.
It is thus natural to ask if genuine quantum effects can help to define new functionality, even if
quantum computing itself is not targeted. This question is especially interesting when interactions
are included, as indeed the latter become pertinent with the confinement of charges caused by the
miniaturisation. Many-body correlations in particular can offer the opportunity to design properties
not achievable through conventional electronics.
In this paper we present such an example in which interactions are tuned
to decouple charge and heat rectification.

\begin{figure}
	\centering
	\includegraphics[width=\columnwidth]{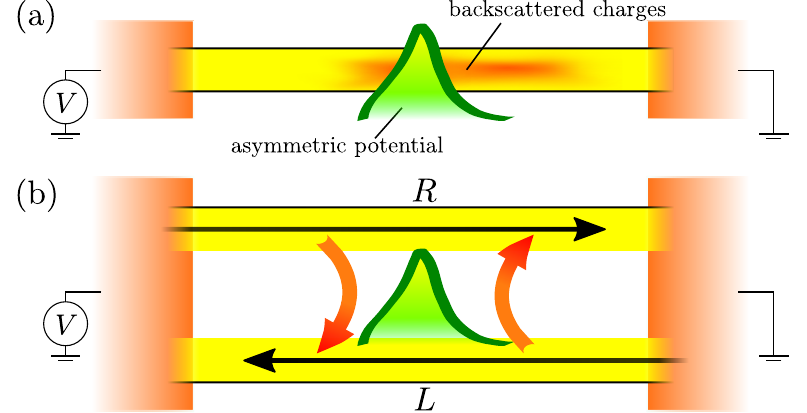}
	\caption{\label{fig:setup}%
	(a) Scheme of the voltage $V$ driven quantum wire with a spatially asymmetric potential $U(x)$.
	Rectification arises from dressing $U(x)$ by backscattered charges together with renormalisation
	through interactions.
	(b) The wire as a thermodynamic system with right (left) moving
	modes $R$ ($L$) in equilibrium with the reservoir on their left (right).
	The reservoirs are fully absorbing for
	incoming particles. Backscattering by $U(x)$ (curved arrows) connects the
	$R$ and $L$ subsystems and causes the transport asymmetry under $V$.
	}
\end{figure}

Rectification, the diode effect, is characterised by an asymmetric current-voltage relation.
In a conventional diode this asymmetry is introduced by $p$ and $n$ type doped sides of a
semiconductor junction. Although the dopants create an electrostatic environment
the resulting physics is understood on the single electron level.
A many-body variant exists but relies on different physics.
It was shown long ago that in quantum wires as illustrated in Fig.\ \ref{fig:setup}
a local scattering potential $U(x)$
causes a strong renormalisation of the current-voltage relation through electron
interactions \cite{Kane1992,Furusaki1993}.
While the leading correction is independent of the potential's form,
sub-leading orders are shape sensitive, and a spatially asymmetric potential induces rectification
\cite{Feldman2005,Braunecker2005,Braunecker2007} which for strong interactions
can become very large.

In this paper we investigate this scenario under the aspect of thermoelectric rectification
where heat current is driven by a voltage $V$. In the nonlinear regime this is different from
a temperature driven current which we do not consider.
The thermoelectric response in quantum wires has been considered in various settings \cite{Krive2001,Li2002,Garg2009,Wakeham2011,DeGottardi2015,Shapiro2017,Ivanov2019}
but for rectification our focus is entirely on the heat flow from backscattering which to our knowledge has not been investigated before.
The many-body setup is also different from the usual approaches to heat rectification that depend on an artful
design of the system or the reservoirs \cite{Li2004,Chang2006,Segal2008,Scheibner2008,Zhan2009,Wu2009,Kobayashi2009,Kuo2010,Roberts2011,Tian2012,Ren2013,Giazotto2013,Martinez-Perez2013,Meair2013,Sanchez2013,Fornieri2014,Jing2015,Martinez-Perez2015,Joulain2016,Ordonez-Miranda2017,Nakai2019}.
With the tools of open quantum systems and quantum thermodynamics we derive an intuitive result that
automatically incorporates the requirement of gauge invariance \cite{Christen1996,Meair2013,Whitney2013,Sanchez2013}
and evaluate it through nonequilibrium perturbation theory.
Remarkably the asymmetry of the heat current appears already at the leading current renormalisation
such that through interactions it decouples from charge rectification and generally dominates.

In addition to normal electrons we consider effectively spinless (e.g.\ polarised or helical) conductors.
For the latter we find that for strong interactions the
heat asymmetry can become as large as the heat current itself whereas the charge asymmetry remains
very small. This produces the phenomenon of a conductor that acts as a heat diode
but not as a charge diode.
Furthermore whether the heat transport is reduced for positive or negative bias
depends on quantum interference and can be switched even through
small changes in the impurity potential which can be created through local external gates.

The plan of the paper is the following: In Sec.\ \ref{sec:phys} we discuss the physics underlying
the heat rectification. The model is introduced in Sec.\ \ref{sec:model} and quantitatively analysed for
noninteracting electrons in Sec.\ \ref{sec:noninteracting}. The modifications from interacting electrons
are the topic of Sec.\ \ref{sec:interacting} in which we also highlight how renormalisation effects cause a decoupling
from charge rectification. Rectification efficiencies are investigated in Sec.\ \ref{sec:efficiency}
before we conclude in Sec.\ \ref{sec:conclusions}. The evaluation of the required correlation functions
relies on standard techniques for one-dimensional conductors and a summary of the used
bosonisation details is provided in the Appendix.


\section{Physics behind rectification}
\label{sec:phys}

For a setup as in Fig.\ \ref{fig:setup} the asymmetry causing rectification is due
to the local potential $U(x)$ alone.
In an interacting system backscattering on $U$, described by Fourier modes $U_{2k_F}$ with $k_F$ the Fermi momentum,
causes a strong renormalisation of transport \cite{Kane1992,Furusaki1993}.
But the usual leading correction, proportional to $|U_{2k_F}|^2$, does not retain spatial information and
does not contribute to rectification.
A spatial asymmetry dependence, which then causes rectification,
appears only at sub-leading orders with higher powers of $U$ in the
renormalisation \cite{Feldman2005,Braunecker2005} but for strong
interactions they can grow in magnitude and indeed create a pronounced diode effect.

In contrast for energy or heat currents the dependence on the asymmetry of $U(x)$ appears remarkably already at leading order.
The potential $U(x)$ contains two contributions, backscattering and forward scattering. The latter does
not affect charge transport but changes locally the kinetic energy of the particles. This is conventionally
captured by a local chemical potential $\mu(x) = \mu + U(x)$, which modifies
the local energy density and enters linearly in the energy transfer by backscattering which remains proportional to $U^2$.
We will show that the overall amplitude is $(U \star U)_{2k_F} U_{2k_F}^*$,
with $\star$ the convolution of the Fourier modes. This amplitude is complex and through its phase
retains the signature of the spatial asymmetry, thus taking the role occurring only at higher orders
for charge rectification. Intuitively this process should be understood as the electrons picking up
a phase due to the locally modified density just before or after backscattering. This phase is different
for counterflowing particles due to the asymmetry of $U(x)$. Since the backscattering process is local
it convolves this phase of the incoming wave packets with the phase from the backscattering potential
to the final momentum transfer $2k_F$ required to change the direction of propagation of wave packets.
The phases of the complex amplitudes are thus the result of the interference of the incoming with the
outgoing wave packets and are sensitive to the precise shape of $U(x)$. Due to the nonlocal nature of the
Fourier transformation there is no intuitive way to anticipate the precise acquired phases but in
Sec.\ \ref{sec:noninteracting} and Fig.\ \ref{fig:alpha} we provide a concrete example
illustrating how external gating can influence the interference to one's advantage.

The dependence of charge and heat current rectification causes then a decoupling of
the corresponding renormalisation channels in an interacting system. This leads
to different voltage $V$ dependences in the form of different power-law scalings.
Since heat rectification arises at the most relevant order it usually dominates over
charge rectification, and interactions can even be tuned such that heat rectification
is strongly enhanced while charge rectification remains very small, both relative
to their total currents.
Such a device then operates as a good thermal diode without significant impact
on charge rectification.


\section{Model and currents}
\label{sec:model}

For a quantitative evaluation we consider a one-dimensional quantum wire connected
to reservoirs on each end. For voltage $V$ driven thermoelectric transport the reservoirs do
not have any specific form and their temperature is irrelevant as long as $V$ sets the dominant
energy scale.
The physics of a one-dimensional conductor is very susceptible to electron interactions,
and the conventional Fermi liquid paradigm is generically replaced by the universality class of the
Luttinger liquid \cite{Haldane1981}. The latter is characterised by collective charge and
spin density modes which in part drastically change the shape of correlation functions in
comparison with the Fermi liquid, and we will make use of this behaviour, in particular of
the property that charge and heat rectification currents are differently renormalised.
Since this physics is universal it can conveniently be accessed through an appropriately chosen
model. We therefore describe the system in terms of
the Tomonaga-Luttinger model \cite{Tomonaga1950,Luttinger1963,Gogolin1998,Giamarchi2007}
in which electron operators $\psi(x)$ are split into right
$R$ and left $L$ moving modes with momenta close to $+k_F$ and $-k_F$.
With $\psi_{R,L}(x)$ the corresponding field operators, the Hamiltonian without
$U(x)$ becomes
\begin{align}
	H &= \int dx \sum_\nu \psi^\dagger_\nu(x) (\mu_\nu- \nu i \hbar v_F \partial_x) \psi_\nu(x)
\notag\\
	&+ \int dx dy \, \mathcal{V}(x-y) \psi^\dagger(x) \psi^\dagger(y) \psi(y) \psi(x),
\label{eq:H}
\end{align}
where $\nu = R,L=+,-$, the integration is over the wire length, $v_F$ is the Fermi velocity
for the linearised dispersion,
and $\mathcal{V}(x-y)$ the (screened) electron interaction. Spin is not written since all
terms are spin-diagonal but its influence is discussed later.
As shown in Fig.\ \ref{fig:setup} (a) the $R$ and $L$ movers have the chemical potentials $\mu_{R,L}$
that are set by the emitting reservoirs \cite{Egger1998,Fisher1997,Krive2001,Feldman2005,Braunecker2005,Braunecker2007}.
The voltage drop is $V = (\mu_R-\mu_L)/e$, with $e$ the electron charge.
For $\mu_R\neq\mu_L$ the Fermi momentum is adjusted to
$k_F^{R,L} = k_F+(\mu_{R,L}-\mu_0)/\hbar v_F$, with $\mu_0$ the equilibrium
chemical potential. The field operator is
$\psi(x) = e^{i k_F^R x} \psi_R(x) + e^{-i k_F^L x} \psi_L(x)$.
Without interactions the Hamiltonian decouples into $R$ and $L$ moving fermionic modes,
providing the condition shown in Fig.\ \ref{fig:setup} (b). Such a
decoupling persists even with interactions if $\pi/k_F$ is not
commensurate with the crystal lattice, and we exclude the latter special cases.
The low energy eigenmodes become then collective density
wave excitations that still are separate $R$ and $L$ movers, albeit both mixing the original $R$ and $L$
movers \cite{Gogolin1998,Giamarchi2007}.
An appropriate proof of this decoupling can be given through the bosonisation technique which allows us
in addition to evaluate all required correlators explicitly. This is a standard calculation which we use
as a tool to supplement the results, but it is not of primary importance for the discussion otherwise.
We therefore relegate the details to the Appendix
and in the main discussion focus entirely on the resulting physics and its interpretation.
Through the decoupling of modes we can thus always write
$H = H_L + H_R$ with $H_\nu$ containing only $\nu$ moving eigenmodes.
Scattering on $U(x)$ has the Hamiltonian
\begin{equation}
	H_U = \int dx \sum_{\nu,\nu'} U(x) e^{-i (\nu k_F^\nu- \nu' k_F^{\nu'}) x} \psi_\nu^\dagger(x) \psi_{\nu'}(x),
\end{equation}
where $U(x)$ is non-zero only in a small region $< \pi/k_F$ around $x=0$ and
we assume that it is spatially asymmetric, $U(x) \neq U(-x)$.
This potential has two roles. For $\nu=\nu'$ it describes
forward scattering that can be added to $H_\nu$ by letting $\mu_\nu(x) = \mu_\nu + U(x)$.
For $\nu\neq \nu'$ the potential introduces backscattering between $R$ and $L$ movers, and we call this part of the Hamiltonian $H_b$.
For a helical system (opposite spins bound to $R,L$ movers) $U$ is a magnetic
impurity inducing both spin preserving forward and spin-flip backward scattering.

Backscattering is a relevant perturbation for electron transport \cite{Kane1992,Furusaki1993} but the leading term, proportional to
$|U_{2k_F}|^2$, is asymmetry insensitive. Thus charge rectification depends on sub-leading contributions \cite{Feldman2005,Braunecker2005},
revealed through the rectification particle current, $\dot{N}^r_\nu=\dot{N}_\nu(V)+\dot{N}_\nu(-V)$, where
$\dot{N}_\nu = \frac{d}{dt}N_{\nu}$
measures how particle numbers $N_\nu$ of $\nu$ movers change by backscattering.
By particle conservation $\dot{N}_R = - \dot{N}_L$.

Identifying heat or energy transfer is a bit more subtle. We have to consider
$R$ and $L$ movers as thermodynamic subsystems that are brought into contact through
the interface Hamiltonian $H_b$ [see Fig.\ \ref{fig:setup} (b)].
The energy flow into system $\nu$ is given by the change of the
internal energy  $E_{\nu} = \mathrm{Tr}_{\nu}\{H_{\nu} \rho_{\nu}\}$,
with $\mathrm{Tr}_{\nu}$ the trace over the degrees of freedom of subsystem $\nu$
and $\rho_{\nu}$ the reduced density matrix obtained from the full density
matrix $\rho$ through $\rho_{R,L} = \mathrm{Tr}_{L,R}\{\rho\}$.
If we put all time dependence in $\rho$ and notice that we can write
$E_\nu = \mathrm{Tr}\{ H_\nu \rho\}$ with $\mathrm{Tr}$ the full trace
we obtain
\begin{equation}
	\dot{E}_\nu
	=
	-\frac{i}{\hbar} \mathrm{Tr} \bigl\{ H_\nu [ H , \rho]\bigr\}
	=
	-\frac{i}{\hbar} \mathrm{Tr} \bigl\{ [ H_\nu , H_b ] \rho \bigr\},
\label{eq:dot_E}
\end{equation}
where we have used the von Neumann equation for the time evolution of $\rho$,
the cyclicity of the trace, and $[H_\nu,H_{\nu'}]=0$.
Notice that $\dot{E}_R = - \dot{E}_L$ although formally they differ by an interface
term $\propto H_b$ \cite{Hossein-Nejad2015}. But in steady state this term has
the expectation value zero.
To identify the heat current $\dot{Q}_\nu$ through the interface $H_b$
we separate $\dot{E}_\nu$ into heat and work fluxes.
The criterion for heat flux as the quantity that changes entropy \cite{Weimer2008,Hossein-Nejad2015}
would give $\dot{Q}_\nu=\dot{E}_\nu$ as $H_b$ mixes $R$ and $L$ states.
But $H_b$ exchanges particles too such that for the grand canonical
setting $\mu_\nu \dot{N}_\nu$ has to be split off from the heat flux and we obtain
$\dot{Q}_\nu = \dot{E}_\nu-\mu_\nu\dot{N}_\nu$. This splitting reproduces the standard form of
the change in thermodynamic potential
$dE_\nu = dQ_\nu + \mu_\nu dN_\nu$
and
is in particular necessary because it makes $\dot{Q}_\nu$ independent of the
gauge fixing the origin of energy \cite{Christen1996,Meair2013,Whitney2013,Sanchez2013}.
Similarly to Eq.\ \eqref{eq:dot_E} we obtain
\begin{align}
	\dot{Q}_\nu
	=
	-\frac{i}{\hbar} \mathrm{Tr} \bigl\{ [ H_\nu -\mu_\nu N_\nu, H_b ] \rho \bigr\}.
\label{eq:dot_Q}
\end{align}
and the $\mu_\nu N_\nu$ term indeed cancels the $\mu_\nu$
in Eq.\ \eqref{eq:H}. The latter equation provides the earlier mentioned intuitive
result for the backscattering induced heat current.


\section{Noninteracting electrons}
\label{sec:noninteracting}

Remarkably heat rectification itself does not need interactions and arises
as a clear quantum interference effect.
Focusing on $\dot{Q}_R$
and using the fact that the $R, L$ decoupling of $H$ can be read off
from the first line in Eq.\ \eqref{eq:H},
the standard anticommutation relations yield
\begin{align}
	&\dot{Q}_R
	=
	\frac{1}{\hbar}\int dx U(x)  e^{i (k_F^L+k_F^R) x}
\notag\\
	&\times
	\mathrm{Tr}\Bigl\{
		\psi_L^\dagger(x)
		\bigl( U(x)-i\hbar v_F \partial_x\bigr) \psi_R(x)
		\rho
	\Bigr\}  + \mathrm{c.c.}
	\label{eq:QR_formal}
\end{align}
Although this result is derived for a noninteracting system we show in the Appendix
that it remains unchanged for an interacting system.
In Eq.\ \eqref{eq:QR_formal} as well as in the interacting case below we can drop the $U$ independent term
as it produces only a logarithmic correction
to the amplitude and no rectification at the considered orders.
Furthermore the $V$ dependence of $k_F^\nu$, in contrast to its role for $\dot{N}_\nu^r$ \cite{Feldman2005,Braunecker2005},
just produces higher powers in $V$ and we set
$k_F^R +k_F^L \approx 2k_F$.
The Keldysh nonequilibrium expansion of $\rho$ in $U$ gives at leading order
\begin{align}
	&\dot{Q}_R
	=
	\frac{-i}{\hbar^2} \int dx  dx' U^2(x) U(x') e^{i 2k_F (x-x')}
	 \int_{-\infty}^0 dt
\notag\\
	&\times
	\bigl\langle
		\bigl[
			\psi_L^\dagger(x,0)
			\psi_R(x,0)
			,
			\psi_R^\dagger(x',t) \psi_L(x',t)
		\bigr]
	\bigr\rangle
	+ \mathrm{c.c.},
\label{eq:QR_org}
\end{align}
where $\psi_\nu$ evolves under $H_\nu$ and the expectation
value is over the uncoupled $R, L$ systems.
Equation \eqref{eq:QR_org} describes the interference of an incoming wave packet with its backscattered
counterpart. Due to the different powers $U^2(x)$ and $U(x')$ and the spatial asymmetry of $U$
this expression breaks the $L$--$R$ symmetry and thus the interference patterns are different for applied $\pm V$
voltages. To obtain a quantitative result for the interference we notice that $\psi_\nu(x,t)$ varies slowly on
the scale $\pi/k_F$ which is much longer than the support of $U(x)$. This allows us to set the
arguments $x,x'$ of the field operators to 0, and the spatial integration then provides
the Fourier transforms $U_k$ of $U(x)$ and $(U^2)_k=(U\star U)_k$ of $U^2(x)$,
\begin{align}
	&\dot{Q}_R
	=
	\frac{-i}{\hbar^2} (U^2)_{2k_F}^* U_{2k_F}
	 \int_{-\infty}^0 dt
\notag\\
	&\times
	\bigl\langle
		\bigl[
			\psi_L^\dagger(0,0)
			\psi_R(0,0)
			,
			\psi_R^\dagger(0,t) \psi_L(0,t)
		\bigr]
	\bigr\rangle
	+ \mathrm{c.c.}
\label{eq:QR}
\end{align}
The gauge transformation $\psi_\nu(x,t) = e^{-i \mu_\nu t/\hbar} \tilde{\psi}_\nu(x,t)$
sets the bulk $\mu_{R,L} \to 0$ but makes the $V$ dependence evident by giving rise to
$e^{i (\mu_R-\mu_L) t/\hbar} = e^{i e V t/\hbar}$ in Eq.\ \eqref{eq:QR}.
If $eV$ is larger than the thermal energy we can neglect temperature for the evaluation of the
correlators, which marks a difference from temperature driven transport \cite{Kane1996}.
The time dependence of the remaining correlators $\langle \tilde{\psi}_\nu^\dagger(0,0) \tilde{\psi}_\nu(0,t) \rangle$
and $\langle \tilde{\psi}_\nu(0,0) \tilde{\psi}_\nu^\dagger(0,t)\rangle$ is then $1/t$,
set by the cutoff of the energy integration by the Fermi surface \cite{Nozieres1969}.
By going to dimensionless variables $y=|eV| t/\hbar$ we see that $\dot{Q}_R$ scales as $|V|$. This
linear response result is expected since Eq.\ \eqref{eq:QR} is identical to
$\dot{N}_R$ except for the $U^2$ amplitude instead of $U$.
If we collect all invariant parameters in the constant $C$ we obtain
\begin{equation}
	\dot{Q}_R
	=
	- (U^2)_{2k_F}^* U_{2k_F}
	C |eV|
	 \int_{-\infty}^0 dy \frac{i \, e^{\mathrm{sign}(V) i y}}{y^2}
	+ \mathrm{c.c.}
\label{eq:Q_R_in_y}
\end{equation}
The divergence at $y \to 0$ in the integral results from the constant density of states in the Tomonoga-Luttinger
model and requires a cutoff by the true bandwidth. This cutoff could in principle produce a further
$V$ dependence from the scaling $t \to y$ but the magnitude of currents is set by $V$ and has
to vanish at $V=0$. Therefore the cutoff must drop out with the commutators in Eq.\ \eqref{eq:QR} and any singularity
can be neglected in the evaluation of the integral.
For $\dot{N}_R$
the first two factors in Eq.\ \eqref{eq:Q_R_in_y} would be $U_{2k_F}^* U_{2k_F} = |U_{2k_F}|^2$
and the expression in front of the integral would be real.
With the `c.c' the integrand then becomes $\mathrm{sign}(V) 2 \sin(y)/y^2$
such that $\dot{N}_R$ just changes sign but not magnitude with $V \to -V$.
Charge rectification thus requires higher order corrections \cite{Feldman2005,Braunecker2005}.

Heat current involves instead $(U^2)_{2k_F}^* U_{2k_F}$.
For a real symmetric potential $U(x)=U(-x)$ the Fourier components are real, and rectification
remains absent. But for a spatially asymmetric potential
$(U^2)_{2k_F}^* U_{2k_F} = |(U^2)_{2k_F} U_{2k_F}| e^{i \alpha}$
is complex with a nonzero phase $\alpha$. The integrand becomes
$2 [\mathrm{sign}(V) \cos(\alpha) \sin(y)+\sin(\alpha) \cos(y)]/y^2$.
The term in $\sin(\alpha)$ is invariant under the sign of $V$ showing that
heat current rectification exists even for a noninteracting system.
If we define $\dot{Q}_\nu^r = \dot{Q}_\nu(V) + \dot{Q}_\nu(-V)$ as the rectification heat current
measuring the asymmetry between $\pm V$ bias, we thus find that
\begin{align}
	\dot{Q}_R^r
	= \sin(\alpha) |V| \ |(U^2)_{2k_F} U_{2k_F}| \ C',
\label{eq:QRr}
\end{align}
where the constant $C'$ absorbs $C$ and the value of the remaining integration.
An identical result holds
for $\dot{Q}^r_L$ with $R \to L$ and $\alpha \to -\alpha$.

\begin{figure}
	\centering
	\includegraphics[width=\columnwidth]{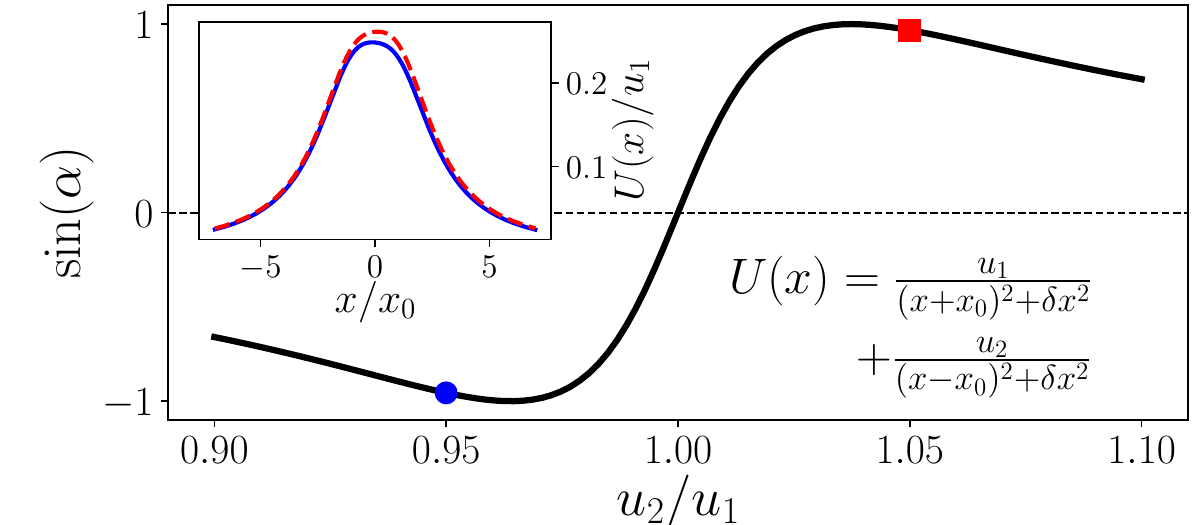}
	\caption{\label{fig:alpha}
	Sensitivity of $\sin(\alpha)$ in Eq.\ \eqref{eq:QRr} to the shape of $U(x)$, given here by the sum
	of two Lorentzians as indicated in the figure with parameters $\delta x=2x_0$
	and $k_F = 0.4/x_0$ in generic units $x_0, u_1$. The inset shows $U(x)$ for the ratios $u_2/u_1$ marked
	by the circle (solid line) and square (dashed line).
	}
\end{figure}

The phase $\alpha$ therefore captures the quantum interference resulting from Eq.\ \eqref{eq:QR_org} and is thus very sensitive to the
precise shape of $U(x)$, such that generally the sign of $\sin(\alpha)$ is arbitrary. But this
sensitivity also allows tuning for which only slight changes of $U(x)$ are required.
In Fig.\ \ref{fig:alpha} we provide an example for $U(x)$ being the sum of two Lorentzians,
and show that even a change in amplitude by just a few percent can completely reverse the polarity.
This sensitivity is a general feature of the asymmetry but independent of the shape of $U(x)$ otherwise.
Asymmetric potentials are naturally realised when impurities appear close together within a Fermi
wavelength \cite{Feldman2005,Braunecker2005,Braunecker2007}. Nearby narrow gates could then ensure sufficient tuneability.
With typical Fermi wavelengths around 100 nm direct creation by state-of-the-art gates could also be considered.


\section{Interacting electrons}
\label{sec:interacting}

Interactions cause a significant renormalisation of the backscattering amplitude and hence of the rectification properties.
This renormalisation occurs because backscattering locally changes the charge density in the vicinity of $U(x)$,
so that an incoming wave packet experiences
the combined effect from the potential $U(x)$ and the interaction with the displaced charges.
This causes a self-consistent dressing of the potential and, for repulsive scatterers, a strong enhancement of
the effective backscattering amplitude \cite{Kane1992,Furusaki1993}.
Underlying this strong response is the fact that in one dimension interactions destabilise the Fermi liquid and cause an instability
towards density fluctuations. The universality class describing this physics is the Luttinger liquid \cite{Haldane1981,Gogolin1998,Giamarchi2007},
and the bosonisation method provides for the latter a standard technique to compute correlation functions
at arbitrary interaction strength. For the present discussion our focus is on the effect of the results obtained through this technique,
and we provide thus the necessary details on how to obtain the results in the Appendix.
The correlators in Eq.\ \eqref{eq:QR}
are then modified from $1/t^2$ to $1/t^\gamma$ \cite{Gogolin1998,Giamarchi2007}
where $\gamma = 2K$ for the spinless case and $\gamma = K_c + K_s$ for the spinful case.
The parameters $K$ and $K_{c,s}$ capture all interactions. $K,K_c=1$ is the noninteracting case,
$0<K,K_c<1$ encodes repulsive interactions, and $K,K_c>1$ encodes attractive interactions.
For the spinful case, if spin SU(2) symmetry is preserved $K_s=1$ and if it is broken $K_s>1$.
We exclude $K_s<1$ as it would represent an instability to spin density waves and require specially tuned spin interactions.

The decoupling into $R$ and $L$ moving eigenmodes persists, and although the eigenmodes turn into renormalised density waves
we show in App.\ \ref{sec:appendix} that Eq.\ \eqref{eq:QR} remains valid.
The evaluation of the correlators shows that they are identical
to those for the backscattering current \cite{Kane1992}
and change the voltage dependence in Eq.\ \eqref{eq:QRr} from $|V|$
to
\begin{equation}
	\dot{Q}_\nu^r \sim |V|^{\gamma-1}.
\end{equation}
Since for repulsive interactions $\gamma<2$ this boosts the rectification current.
In comparison charge rectification $\dot{N}_\nu^r = \dot{N}_\nu(V)+\dot{N}_\nu(-V)$
scales with $|V|^{\gamma_c}$ where
$\gamma_c = \min(2K,6K-2)$ for spinless electrons \cite{Feldman2005} and
$\gamma_c = \min(K_c+K_s,4K_c,3K_c+K_s-2, 12 K_c-2)$ for spinful electrons \cite{Braunecker2005}.
Heat and charge rectification thus decouple, and since $\dot{Q}_\nu^r$
arises from higher relevant contributions it is usually more significant.

\begin{figure}
	\centering
	\includegraphics[width=\columnwidth]{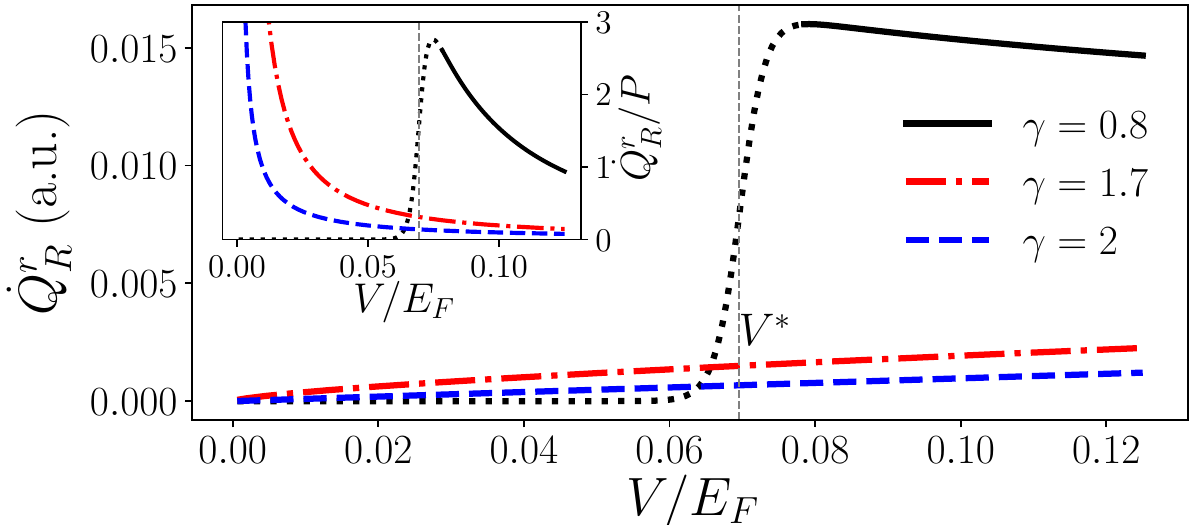}
	\caption{\label{fig:QV}%
	Heat rectification as a function of voltage for $U$ corresponding to the red square in Fig.\ \ref{fig:alpha}
	(with $u_1/E_F=0.7$ and $E_F$ setting the order of the bandwidth).
	Interactions with $\gamma<2$
	enhance the noninteracting $\gamma=2$. For $\gamma < 1$ (possible only for
	spin polarised electrons) a maximum enhancement is reached near $V=V^*$ where Eq.\ \eqref{eq:QRr}
	crosses over to strong coupling scaling and $\dot{Q}_R^r$ decreases again to zero
	(expected trend shown by the dotted line). The inset shows the corresponding efficiency $\dot{Q}_R^r/P$,
	with $P$ the dissipated power.
	While the scaling is exact only the order of magnitude is known for the amplitudes
	and we have set $C'=1$.
	}
\end{figure}

Notice that these currents are obtained perturbatively on top of the heat or charge transfer between the
reservoirs which from standard transport theory are proportional to $V$.
Particularly interesting is when $\gamma-1$ or $\gamma_c$ becomes negative. Then the
current increases when lowering $V$ until at some $V^*$ it becomes as large as the unperturbed current
$\propto V$.
Perturbation theory must then be replaced by a strong coupling calculation.
Since currents must vanish at $V=0$ the currents decrease then again.
Near $V^*$ backscattering and thus rectification
are largest. Since $K_s \geq 1$ the spinful case never has $\gamma < 1$ but this
can be achieved for spin polarised electrons when $K<1/2$. If furthermore $K>1/3$
then charge rectification keeps $\gamma_c>0$  \cite{Feldman2005}, making the decoupling of heat and charge
rectification most pronounced, with strongly rectified heat and only weakly
asymmetric charge current. Figure \ref{fig:QV} shows $\dot{Q}_R^r$ for different $\gamma$.
For $\gamma < 1$ we interpolate to the strong coupling scaling
$\dot{Q}_R^r \sim |V|^{4/\gamma-1}$ \cite{Kane1992} across $V^*$.


\section{Rectification efficiency}
\label{sec:efficiency}

For a good diode the ratio $r=\dot{Q}_R(-V)/\dot{Q}_R(V)$ is either $r \ll 1$ or $r \gg 1$.
In the Tomonaga-Luttinger model an exact calculation of $r$ is tricky due to the required cutoffs.
But Eq.\ \eqref{eq:Q_R_in_y} shows that
\begin{equation}
	r = \frac{A \sin(\alpha) - B \cos(\alpha)}{A \sin(\alpha) + B \cos(\alpha)},
\end{equation}
where $A$ and $B$ are of the same order.
Therefore $r$ is tuneable through $\alpha$ to any value. Although its initial
value is arbitrary this provides the advantage by tuning through gates. In Sec.\ \ref{sec:noninteracting}
we indeed highlighted the sensitivity of $\alpha$ to small changes of gating bias such that
the nonuniversality of $r$ can be used to turn the system into an actively programmable heat diode.

The efficiency of the heat transport is assessed by comparing $\dot{Q}_R^r$ to
the total dissipated power $P = I V$ (Joule heating). Since the total current $I \propto V$
we obtain $\dot{Q}_R^r/P \sim |V|^{\gamma-3}$.
For $1<\gamma<2$ the divergence at $V\to 0$ indicates that heat rectification is most effective when
dissipation is generally low.
For $\gamma < 1$ there is a strong suppression at $V < V^*$ and
the benefit of strong rectification near $V^*$ involves a larger dissipation. This behaviour is
illustrated in the inset of Fig.\ \ref{fig:QV}. Notice that
since the temperature of the reservoirs does not appear in these considerations there is no counterpart
of the thermoelectric figure of merit $ZT$ and we use $\dot{Q}_R^r/P$ instead.


\section{Conclusions}
\label{sec:conclusions}

We have investigated how asymmetric potentials cause heat current rectification in quantum wires.
Although the effect appears already for noninteracting particles through interference of the
backscattered wave packets, it becomes most interesting in an interacting system. For the latter
the charge and heat rectification decouple and are characterised by a different voltage dependence.
The decoupling becomes more pronounced with increasing interactions, and in particular for
spin polarised electrons at interaction strengths $1/3<K<1/2$ we predict a strong effect in which
the heat rectification is strong but charge rectification remains weak. Such interaction strengths
are not untypical for high quality conductors.
For instance, GaAs quantum wires can be tuned to $K_c \sim 0.4$ \cite{Auslaender2002,Steinberg2008}
and are candidates for a helical transition \cite{Scheller2014} that would provide the further
reduction of the spin degree of freedom.
We have furthermore discussed that the rectification polarity is easily manipulatable through local gating.
The basis of this is the sensitive dependence of the quantum interference amplitude on the detailed shape
of the backscattering potential. This sensitivity makes the amplitude nonuniversal, which is quite common
for one-dimensional systems, but it makes it thus also very suitable for an easy tuning of the polarisation
of the rectifier. We have indeed shown that changes of
a few percent of the asymmetric shape of the potential can be sufficient for a full polarisation reversal.
This can be achieved through local gating and hence could make such a system useful as a programmable
heat rectifier.


\begin{acknowledgments}
We thank P.\ Jacquod for a discussion that strongly inspired this work,
and we thank D.\ E.\ Feldman, P.\ Jacquod, J.\ B.\ Marston, and Z.\ Zhuang for helpful comments.
C.S.\ acknowledges the support from the EPSRC under Grant No. EP/N509759/1.
The work presented in this paper is theoretical. No data was
produced and supporting research data is not required.
\end{acknowledgments}

\appendix

\section{Energy currents for interacting systems in bosonisation formalism}
\label{sec:appendix}

The analysis of backscattering induced heat rectification in the main text relies on the treatment of $R$ and $L$ moving modes
as two distinguishable transport channels. The decoupling of these two channels appears naturally for noninteracting systems
but becomes more subtle when interactions are involved. Indeed the general form of the interaction Hamiltonian
\begin{equation}
	H_\text{int} = \int dx dy \, \mathcal{V}(x-y) \psi^\dagger(x) \psi^\dagger(y) \psi(y) \psi(x),
\end{equation}
with $\psi(x) = e^{i k_F^R x} \psi_R(x) + e^{-i k_F^L x} \psi_L(x)$,
clearly couples $R$ and $L$ movers. However, as long as the fermion density is not commensurate with the underlying lattice
most of the terms in $H_\text{int}$ are irrelevant in the renormalisation group sense \cite{Gogolin1998,Giamarchi2007},
and the only remaining interactions are of the form
\begin{equation}  \label{eq:H_int}
	H_\text{int} = \sum_{\nu,\nu'=R,L} \int dx dy \, \mathcal{V}(x-y) \psi_\nu^\dagger(x) \psi_{\nu'}^\dagger(y) \psi_{\nu'}(y) \psi_\nu(x).
\end{equation}
Within the Luttinger liquid paradigm (see \cite{Gogolin1998,Giamarchi2007} for full details on the formalism used in this appendix)
the fermionic Hamiltonian is then mapped onto a set of bosonic, harmonic oscillator type Hamiltonians,
with boson fields representing the density fluctuations of the $R$ and $L$ movers. The interactions in Eq.\ \eqref{eq:H_int}
cause a coupling between the $R$ and $L$ type boson fields for $\nu \neq \nu'$, but this coupling remains bilinear so that
the Hamiltonian is a quadratic form described by a $2 \times 2$ matrix for the $R, L$ fields which can be straightforwardly
diagonalised. Although they mix contributions from both the original $R$ and $L$ movers, the resulting
eigenmodes $\phi_{R,L}$ describe wave packets that move only to the right or the left, and hence maintain effectively the decoupling of
$R$ and $L$ moving modes.
The resulting Hamiltonians can be written for the spinless (or spin polarised) case as
\begin{equation} \label{eq:H_nu}
	H_\nu = \int dx \, \frac{v}{4\pi K} \, \bigl(\partial_x \phi_\nu(x)\bigr)^2 + \mu_\nu N_\nu,
\end{equation}
for $\nu = R,L$, where the fields obey the commutation relations
\begin{equation} \label{eq:comm}
	[\phi_\nu(x'),\partial_x\phi_{\nu'}(x)] = i \pi K \delta_{\nu,\nu'} \delta(x-x'),
\end{equation}
that is \ $\phi_\nu$ and $\partial_x \phi_\nu$ are conjugate boson fields up to a normalisation.
The parameter $K$ results from the diagonalisation of the $2\times 2$ matrix and thus encodes the entire effect
of $H_\text{int}$. It takes the values discussed in Sec.\ \ref{sec:interacting}.

The term $\mu_\nu N_\nu$ in Eq.\ \eqref{eq:H_nu} contains the energy correction from the chemical potentials $\mu_\nu$ and
the particle numbers $N_\nu$ of $\nu$ movers. This term depends on the choice of gauge for $\mu_\nu$
but drops out in the gauge independent expressions considered below and for the heat currents $\dot{Q}_\nu$
discussed in the main text.

The original fermion operators are expressed in terms of these eigenmodes as
\begin{equation}  \label{eq:psi}
	\psi_\nu(x) = \frac{\eta_\nu}{\sqrt{2\pi a}} e^{-\frac{i}{2}(\nu-K^{-1})\phi_L(x) -\frac{i}{2} (\nu+K^{-1})\phi_R(x)},
\end{equation}
with the signs $\nu=R=+$ and $\nu=L=-$, and $a$ a short distance cutoff, typically on the order of the lattice spacing.
The $\eta_\nu$ are operators that lower the overall fermion number by 1 and
guarantee the fermionic exchange statistics. For the further analysis they do not play any further role and
can be dropped henceforth.

The forward scattering term on the impurity is obtained from point splitting of the densities,
\begin{equation}
	\psi_\nu^\dagger(x)\psi_\nu(x) =
	\frac{\nu+K^{-1}}{2\pi}\partial_x\phi_R(x)
	+
	\frac{\nu-K^{-1}}{2\pi}\partial_x\phi_L(x).
\end{equation}
We have omitted here a term proportional to the average particle density $k_{F,\nu}$
as it contributes only a constant to the Hamiltonian.
Consequently we have
\begin{equation}
	U(x)\sum_\nu \psi_\nu^\dagger(x) \psi_\nu(x)
	= \frac{U(x)}{\pi K} \bigl(\partial_x \phi_R(x) - \partial_x \phi_L(x) \bigr).
\end{equation}
This term thus separates well into $R$ and $L$ moving contributions such that the total $\nu$
moving Hamiltonian after subtraction of the offset by the chemical potential reads
\begin{equation}
	H_\nu -\mu_\nu N_\nu
	= \int dx \Bigl[
		\frac{v}{4\pi K} \, \bigl(\partial_x \phi_\nu(x)\bigr)^2
		+
		\nu \frac{U(x)}{\pi K} \partial_x \phi_\nu(x)
	\Bigr].
\end{equation}
On the other hand, the backscattering Hamiltonian $H_b$ becomes
\begin{align}
	H_b
	&=
	\int dx \, U(x)
	e^{-2 i k_F x}
	\psi_R^\dagger(x) \psi_L(x) + \text{h.c.}
\nonumber\\
	&=
	\int dx \,
	\frac{U(x)}{2\pi a}
	e^{-2 i k_F x}
	e^{i\phi_L(x) + i \phi_R(x)}
	+ \text{h.c.},
\label{eq:Hb}
\end{align}
with $2k_F = k_F^R + k_F^L$. The interaction caused renormalisation causes further
effective multi-particle backscattering terms in $H_b$ \cite{Kane1992,Furusaki1993}.
For the main correction to the current and thus the main contribution to $\dot{Q}_\nu^r$ these
are less relevant though and do not need to be considered. Charge rectification on the other hand
depends directly on these multi-particle terms which were accordingly analysed in detail
in Refs.\ \cite{Feldman2005,Braunecker2005,Braunecker2007}, hence the cited different scaling
laws for charge rectification.

For the heat transferred through backscattering, expressed by the commutator
$[H_\nu-\mu_\nu N_\nu, H_b]$ we therefore need to evaluate the commutators of
$(\partial_x \phi_\nu)^2$ and $\partial_x \phi_\nu$ with $H_b$ as given in Eq.\ \eqref{eq:Hb}. From the
commutation relations \eqref{eq:comm} we see that
\begin{equation}
	\bigl[ \partial_x \phi_\nu(x), e^{i \phi_L(x') + i \phi_R(x')} \bigr]
	=
	\pi K \, e^{i \phi_L(x) + i \phi_R(x)} \delta(x-x').
\end{equation}
The commutator with $(\partial_x \phi_\nu)^2$ takes a similar form with an additional
factor $\partial_x \phi_\nu$. As noted the main text such terms produce less relevant
logarithmic corrections to the leading expression so that we can leave them aside.
The commutator for the heat backscattering current is then given by
\begin{align}
	&[H_\nu-\mu_\nu N_\nu, H_b]
\notag\\
	&=
	\nu \int dx \, U^2(x) \frac{e^{-2i k_F x}}{2\pi a} e^{i \phi_L(x) + i \phi_R(x)} - \text{h.c.}
\notag\\
	&=
	\nu \int dx \, U^2(x) e^{-2i k_F x} \psi_R^\dagger(x) \psi_L(x) - \text{h.c.}
\label{eq:heat_curr}
\end{align}
This is exactly the same result obtained from the pure fermionic description of
Eq.\ \eqref{eq:QR_formal} but obtained here for a general interacting system with arbitrary values of $K$.
It is notable that the backscattering heat current operator has no direct $K$ dependence.

When the spin degree of freedom is taken into account the bosonic fields double
into charge and spin fluctuations but the structure of the equations and the identities are identical,
up to extra charge and spin labels. Equation \eqref{eq:heat_curr}
is again unchanged from its noninteracting fermionic expression.

Finally the evaluation of the correlation functions in Eq.\ \eqref{eq:QR} follows the
standard method.
The expectation values in Eq.\ \eqref{eq:QR} can be reduced to the computation
of bosonic correlators through the identity
$\langle e^{i\phi_\nu(t)} e^{-i\phi_\nu(0)} \rangle = e^{\langle \phi_\nu(t) \phi_\nu(0) - [\phi_\nu^2(t)+\phi^2_\nu(0)]/2\rangle}$
which is valid for a quadratic bosonic theory. Here we have set $\phi_\nu(t) = \phi_\nu(x=0,t)$.
Focusing again on the spinless case the bosonic correlators are then evaluated as \cite{Gogolin1998,Giamarchi2007}
\begin{equation} \label{eq:corr}
	\langle \phi_\nu(t) \phi_\nu(0) - [\phi_\nu^2(t)+\phi_\nu^2(0)]/2 \rangle
	= -K \ln\left[(ia - v t)/ia\right],
\end{equation}
where $a$ is the short distance cutoff of the theory and $v=v_F/K$ the interaction renormalised
Fermi velocity. The correlators in Eq.\ \eqref{eq:QR} lead to the exponential of two such bosonic
correlators which thus provides the time dependence $1/t^\gamma$ with $\gamma = 2K$ as discussed
in Sec.\ \ref{sec:interacting}. The limit $K=1$ matches indeed the noninteracting case.

For the spinful case
the eigenmodes $\phi_{\lambda,\nu}$ acquire the further index $\lambda=c,s$ expressing the
charge and spin degrees of freedom.
The latter are independent and obey the same commutation relations \eqref{eq:comm} with
an additional $\delta_{\lambda,\lambda'}$ factor. The Hamiltonian decomposes into four terms $H_{\lambda,\nu}$
each of which is of the form of Eq.\ \eqref{eq:H_nu} with the replacement $K \to K_\lambda$.
The same replacement of $K$ is made for the correlators in Eq.\ \eqref{eq:corr}.
For fermion operators the exponential in Eq.\ \eqref{eq:psi} is replaced by
$\psi_{\nu,\sigma} \sim e^{-i 2^{-3/2}(\varphi_{c,\nu}+\sigma \varphi_{s,\nu})}$
where $\sigma = \pm$ is the additional spin index and
$\varphi_{\lambda,\nu} = (\nu-K_\lambda^{-1}) \phi_{\lambda,L} + (\nu+K_\lambda^{-1}) \phi_{\lambda,R}$.
Consequently the exponent $\gamma$ for the time dependence in the heat current correlators
is replaced by $\gamma = K_c + K_s$ as described in Sec.\ \ref{sec:interacting}.

\




\begin{thebibliography}{49}%
\makeatletter
\providecommand \@ifxundefined [1]{%
 \@ifx{#1\undefined}
}%
\providecommand \@ifnum [1]{%
 \ifnum #1\expandafter \@firstoftwo
 \else \expandafter \@secondoftwo
 \fi
}%
\providecommand \@ifx [1]{%
 \ifx #1\expandafter \@firstoftwo
 \else \expandafter \@secondoftwo
 \fi
}%
\providecommand \natexlab [1]{#1}%
\providecommand \enquote  [1]{``#1''}%
\providecommand \bibnamefont  [1]{#1}%
\providecommand \bibfnamefont [1]{#1}%
\providecommand \citenamefont [1]{#1}%
\providecommand \href@noop [0]{\@secondoftwo}%
\providecommand \href [0]{\begingroup \@sanitize@url \@href}%
\providecommand \@href[1]{\@@startlink{#1}\@@href}%
\providecommand \@@href[1]{\endgroup#1\@@endlink}%
\providecommand \@sanitize@url [0]{\catcode `\\12\catcode `\$12\catcode
  `\&12\catcode `\#12\catcode `\^12\catcode `\_12\catcode `\%12\relax}%
\providecommand \@@startlink[1]{}%
\providecommand \@@endlink[0]{}%
\providecommand \url  [0]{\begingroup\@sanitize@url \@url }%
\providecommand \@url [1]{\endgroup\@href {#1}{\urlprefix }}%
\providecommand \urlprefix  [0]{URL }%
\providecommand \Eprint [0]{\href }%
\providecommand \doibase [0]{http://dx.doi.org/}%
\providecommand \selectlanguage [0]{\@gobble}%
\providecommand \bibinfo  [0]{\@secondoftwo}%
\providecommand \bibfield  [0]{\@secondoftwo}%
\providecommand \translation [1]{[#1]}%
\providecommand \BibitemOpen [0]{}%
\providecommand \bibitemStop [0]{}%
\providecommand \bibitemNoStop [0]{.\EOS\space}%
\providecommand \EOS [0]{\spacefactor3000\relax}%
\providecommand \BibitemShut  [1]{\csname bibitem#1\endcsname}%
\let\auto@bib@innerbib\@empty
\bibitem [{\citenamefont {Kane}\ and\ \citenamefont {Fisher}(1992)}]{Kane1992}%
  \BibitemOpen
  \bibfield  {author} {\bibinfo {author} {\bibfnamefont {C.~L.}\ \bibnamefont
  {Kane}}\ and\ \bibinfo {author} {\bibfnamefont {M.~P.~A.}\ \bibnamefont
  {Fisher}},\ }\href {\doibase 10.1103/PhysRevB.46.15233} {\bibfield  {journal}
  {\bibinfo  {journal} {Phys. Rev. B}\ }\textbf {\bibinfo {volume} {46}},\
  \bibinfo {pages} {15233} (\bibinfo {year} {1992})}\BibitemShut {NoStop}%
\bibitem [{\citenamefont {Furusaki}\ and\ \citenamefont
  {Nagaosa}(1993)}]{Furusaki1993}%
  \BibitemOpen
  \bibfield  {author} {\bibinfo {author} {\bibfnamefont {A.}~\bibnamefont
  {Furusaki}}\ and\ \bibinfo {author} {\bibfnamefont {N.}~\bibnamefont
  {Nagaosa}},\ }\href {\doibase 10.1103/PhysRevB.47.4631} {\bibfield  {journal}
  {\bibinfo  {journal} {Phys. Rev. B}\ }\textbf {\bibinfo {volume} {47}},\
  \bibinfo {pages} {4631} (\bibinfo {year} {1993})}\BibitemShut {NoStop}%
\bibitem [{\citenamefont {Feldman}\ \emph {et~al.}(2005)\citenamefont
  {Feldman}, \citenamefont {Scheidl},\ and\ \citenamefont
  {Vinokur}}]{Feldman2005}%
  \BibitemOpen
  \bibfield  {author} {\bibinfo {author} {\bibfnamefont {D.~E.}\ \bibnamefont
  {Feldman}}, \bibinfo {author} {\bibfnamefont {S.}~\bibnamefont {Scheidl}}, \
  and\ \bibinfo {author} {\bibfnamefont {V.~M.}\ \bibnamefont {Vinokur}},\
  }\href {\doibase 10.1103/PhysRevLett.94.186809} {\bibfield  {journal}
  {\bibinfo  {journal} {Phys. Rev. Lett.}\ }\textbf {\bibinfo {volume} {94}},\
  \bibinfo {pages} {186809} (\bibinfo {year} {2005})}\BibitemShut {NoStop}%
\bibitem [{\citenamefont {Braunecker}\ \emph {et~al.}(2005)\citenamefont
  {Braunecker}, \citenamefont {Feldman},\ and\ \citenamefont
  {Marston}}]{Braunecker2005}%
  \BibitemOpen
  \bibfield  {author} {\bibinfo {author} {\bibfnamefont {B.}~\bibnamefont
  {Braunecker}}, \bibinfo {author} {\bibfnamefont {D.~E.}\ \bibnamefont
  {Feldman}}, \ and\ \bibinfo {author} {\bibfnamefont {J.~B.}\ \bibnamefont
  {Marston}},\ }\href {\doibase 10.1103/PhysRevB.72.125311} {\bibfield
  {journal} {\bibinfo  {journal} {Phys. Rev. B}\ }\textbf {\bibinfo {volume}
  {72}},\ \bibinfo {pages} {125311} (\bibinfo {year} {2005})}\BibitemShut
  {NoStop}%
\bibitem [{\citenamefont {Braunecker}\ \emph {et~al.}(2007)\citenamefont
  {Braunecker}, \citenamefont {Feldman},\ and\ \citenamefont
  {Li}}]{Braunecker2007}%
  \BibitemOpen
  \bibfield  {author} {\bibinfo {author} {\bibfnamefont {B.}~\bibnamefont
  {Braunecker}}, \bibinfo {author} {\bibfnamefont {D.~E.}\ \bibnamefont
  {Feldman}}, \ and\ \bibinfo {author} {\bibfnamefont {F.}~\bibnamefont {Li}},\
  }\href {\doibase 10.1103/PhysRevB.76.085119} {\bibfield  {journal} {\bibinfo
  {journal} {Phys. Rev. B}\ }\textbf {\bibinfo {volume} {76}},\ \bibinfo
  {pages} {085119} (\bibinfo {year} {2007})}\BibitemShut {NoStop}%
\bibitem [{\citenamefont {Krive}\ \emph {et~al.}(2001)\citenamefont {Krive},
  \citenamefont {Romanovsky}, \citenamefont {Bogachek}, \citenamefont
  {Scherbakov},\ and\ \citenamefont {Landman}}]{Krive2001}%
  \BibitemOpen
  \bibfield  {author} {\bibinfo {author} {\bibfnamefont {I.~V.}\ \bibnamefont
  {Krive}}, \bibinfo {author} {\bibfnamefont {I.~A.}\ \bibnamefont
  {Romanovsky}}, \bibinfo {author} {\bibfnamefont {E.~N.}\ \bibnamefont
  {Bogachek}}, \bibinfo {author} {\bibfnamefont {A.~G.}\ \bibnamefont
  {Scherbakov}}, \ and\ \bibinfo {author} {\bibfnamefont {U.}~\bibnamefont
  {Landman}},\ }\href {\doibase 10.1063/1.1414571} {\bibfield  {journal}
  {\bibinfo  {journal} {Low Temp. Phys.}\ }\textbf {\bibinfo {volume} {27}},\
  \bibinfo {pages} {821} (\bibinfo {year} {2001})}\BibitemShut {NoStop}%
\bibitem [{\citenamefont {Li}\ and\ \citenamefont {Orignac}(2002)}]{Li2002}%
  \BibitemOpen
  \bibfield  {author} {\bibinfo {author} {\bibfnamefont {M.~R.}\ \bibnamefont
  {Li}}\ and\ \bibinfo {author} {\bibfnamefont {E.}~\bibnamefont {Orignac}},\
  }\href {\doibase 10.1209/epl/i2002-00282-0} {\bibfield  {journal} {\bibinfo
  {journal} {Europhys. Lett.}\ }\textbf {\bibinfo {volume} {60}},\ \bibinfo
  {pages} {432} (\bibinfo {year} {2002})}\BibitemShut {NoStop}%
\bibitem [{\citenamefont {Garg}\ \emph {et~al.}(2009)\citenamefont {Garg},
  \citenamefont {Rasch}, \citenamefont {Shimshoni},\ and\ \citenamefont
  {Rosch}}]{Garg2009}%
  \BibitemOpen
  \bibfield  {author} {\bibinfo {author} {\bibfnamefont {A.}~\bibnamefont
  {Garg}}, \bibinfo {author} {\bibfnamefont {D.}~\bibnamefont {Rasch}},
  \bibinfo {author} {\bibfnamefont {E.}~\bibnamefont {Shimshoni}}, \ and\
  \bibinfo {author} {\bibfnamefont {A.}~\bibnamefont {Rosch}},\ }\href
  {\doibase 10.1103/PhysRevLett.103.096402} {\bibfield  {journal} {\bibinfo
  {journal} {Phys. Rev. Lett.}\ }\textbf {\bibinfo {volume} {103}},\ \bibinfo
  {pages} {096402} (\bibinfo {year} {2009})}\BibitemShut {NoStop}%
\bibitem [{\citenamefont {Wakeham}\ \emph {et~al.}(2011)\citenamefont
  {Wakeham}, \citenamefont {Bangura}, \citenamefont {Xu}, \citenamefont
  {Mercure}, \citenamefont {Greenblatt},\ and\ \citenamefont
  {Hussey}}]{Wakeham2011}%
  \BibitemOpen
  \bibfield  {author} {\bibinfo {author} {\bibfnamefont {N.}~\bibnamefont
  {Wakeham}}, \bibinfo {author} {\bibfnamefont {A.~F.}\ \bibnamefont
  {Bangura}}, \bibinfo {author} {\bibfnamefont {X.}~\bibnamefont {Xu}},
  \bibinfo {author} {\bibfnamefont {J.~F.}\ \bibnamefont {Mercure}}, \bibinfo
  {author} {\bibfnamefont {M.}~\bibnamefont {Greenblatt}}, \ and\ \bibinfo
  {author} {\bibfnamefont {N.~E.}\ \bibnamefont {Hussey}},\ }\href {\doibase
  10.1038/ncomms1406} {\bibfield  {journal} {\bibinfo  {journal} {Nat.
  Commun.}\ }\textbf {\bibinfo {volume} {2}},\ \bibinfo {pages} {396} (\bibinfo
  {year} {2011})}\BibitemShut {NoStop}%
\bibitem [{\citenamefont {DeGottardi}\ and\ \citenamefont
  {Matveev}(2015)}]{DeGottardi2015}%
  \BibitemOpen
  \bibfield  {author} {\bibinfo {author} {\bibfnamefont {W.}~\bibnamefont
  {DeGottardi}}\ and\ \bibinfo {author} {\bibfnamefont {K.~A.}\ \bibnamefont
  {Matveev}},\ }\href {\doibase 10.1103/PhysRevLett.114.236405} {\bibfield
  {journal} {\bibinfo  {journal} {Phys. Rev. Lett.}\ }\textbf {\bibinfo
  {volume} {114}},\ \bibinfo {pages} {236405} (\bibinfo {year}
  {2015})}\BibitemShut {NoStop}%
\bibitem [{\citenamefont {Shapiro}\ \emph {et~al.}(2017)\citenamefont
  {Shapiro}, \citenamefont {Feldman}, \citenamefont {Mirlin},\ and\
  \citenamefont {Shnirman}}]{Shapiro2017}%
  \BibitemOpen
  \bibfield  {author} {\bibinfo {author} {\bibfnamefont {D.~S.}\ \bibnamefont
  {Shapiro}}, \bibinfo {author} {\bibfnamefont {D.~E.}\ \bibnamefont
  {Feldman}}, \bibinfo {author} {\bibfnamefont {A.~D.}\ \bibnamefont {Mirlin}},
  \ and\ \bibinfo {author} {\bibfnamefont {A.}~\bibnamefont {Shnirman}},\
  }\href {\doibase 10.1103/PhysRevB.95.195425} {\bibfield  {journal} {\bibinfo
  {journal} {Phys. Rev. B}\ }\textbf {\bibinfo {volume} {95}},\ \bibinfo
  {pages} {195425} (\bibinfo {year} {2017})}\BibitemShut {NoStop}%
\bibitem [{\citenamefont {Ivanov}\ and\ \citenamefont
  {Uryupin}(2019)}]{Ivanov2019}%
  \BibitemOpen
  \bibfield  {author} {\bibinfo {author} {\bibfnamefont {Y.~V.}\ \bibnamefont
  {Ivanov}}\ and\ \bibinfo {author} {\bibfnamefont {O.~N.}\ \bibnamefont
  {Uryupin}},\ }\href {\doibase 10.1134/S1063782619050075} {\bibfield
  {journal} {\bibinfo  {journal} {Semic.}\ }\textbf {\bibinfo {volume} {53}},\
  \bibinfo {pages} {641} (\bibinfo {year} {2019})}\BibitemShut {NoStop}%
\bibitem [{\citenamefont {Li}\ \emph {et~al.}(2004)\citenamefont {Li},
  \citenamefont {Wang},\ and\ \citenamefont {Casati}}]{Li2004}%
  \BibitemOpen
  \bibfield  {author} {\bibinfo {author} {\bibfnamefont {B.}~\bibnamefont
  {Li}}, \bibinfo {author} {\bibfnamefont {L.}~\bibnamefont {Wang}}, \ and\
  \bibinfo {author} {\bibfnamefont {G.}~\bibnamefont {Casati}},\ }\href
  {\doibase 10.1103/PhysRevLett.93.184301} {\bibfield  {journal} {\bibinfo
  {journal} {Phys. Rev. Lett.}\ }\textbf {\bibinfo {volume} {93}},\ \bibinfo
  {pages} {184301} (\bibinfo {year} {2004})}\BibitemShut {NoStop}%
\bibitem [{\citenamefont {Chang}\ \emph {et~al.}(2006)\citenamefont {Chang},
  \citenamefont {Okawa}, \citenamefont {Majumdar},\ and\ \citenamefont
  {Zettl}}]{Chang2006}%
  \BibitemOpen
  \bibfield  {author} {\bibinfo {author} {\bibfnamefont {C.~W.}\ \bibnamefont
  {Chang}}, \bibinfo {author} {\bibfnamefont {D.}~\bibnamefont {Okawa}},
  \bibinfo {author} {\bibfnamefont {A.}~\bibnamefont {Majumdar}}, \ and\
  \bibinfo {author} {\bibfnamefont {A.}~\bibnamefont {Zettl}},\ }\href
  {\doibase 10.1126/science.1132898} {\bibfield  {journal} {\bibinfo  {journal}
  {Science}\ }\textbf {\bibinfo {volume} {314}},\ \bibinfo {pages} {1121}
  (\bibinfo {year} {2006})}\BibitemShut {NoStop}%
\bibitem [{\citenamefont {Segal}(2008)}]{Segal2008}%
  \BibitemOpen
  \bibfield  {author} {\bibinfo {author} {\bibfnamefont {D.}~\bibnamefont
  {Segal}},\ }\href {\doibase 10.1103/PhysRevLett.100.105901} {\bibfield
  {journal} {\bibinfo  {journal} {Phys. Rev. Lett.}\ }\textbf {\bibinfo
  {volume} {100}},\ \bibinfo {pages} {105901} (\bibinfo {year}
  {2008})}\BibitemShut {NoStop}%
\bibitem [{\citenamefont {Scheibner}\ \emph {et~al.}(2008)\citenamefont
  {Scheibner}, \citenamefont {K{\"{o}}nig}, \citenamefont {Reuter},
  \citenamefont {Wieck}, \citenamefont {Gould}, \citenamefont {Buhmann},\ and\
  \citenamefont {Molenkamp}}]{Scheibner2008}%
  \BibitemOpen
  \bibfield  {author} {\bibinfo {author} {\bibfnamefont {R.}~\bibnamefont
  {Scheibner}}, \bibinfo {author} {\bibfnamefont {M.}~\bibnamefont
  {K{\"{o}}nig}}, \bibinfo {author} {\bibfnamefont {D.}~\bibnamefont {Reuter}},
  \bibinfo {author} {\bibfnamefont {A.~D.}\ \bibnamefont {Wieck}}, \bibinfo
  {author} {\bibfnamefont {C.}~\bibnamefont {Gould}}, \bibinfo {author}
  {\bibfnamefont {H.}~\bibnamefont {Buhmann}}, \ and\ \bibinfo {author}
  {\bibfnamefont {L.~W.}\ \bibnamefont {Molenkamp}},\ }\href {\doibase
  10.1088/1367-2630/10/8/083016} {\bibfield  {journal} {\bibinfo  {journal}
  {New J. Phys.}\ }\textbf {\bibinfo {volume} {10}},\ \bibinfo {pages} {083016}
  (\bibinfo {year} {2008})}\BibitemShut {NoStop}%
\bibitem [{\citenamefont {Zhan}\ \emph {et~al.}(2009)\citenamefont {Zhan},
  \citenamefont {Li}, \citenamefont {Kohler},\ and\ \citenamefont
  {H{\"{a}}nggi}}]{Zhan2009}%
  \BibitemOpen
  \bibfield  {author} {\bibinfo {author} {\bibfnamefont {F.}~\bibnamefont
  {Zhan}}, \bibinfo {author} {\bibfnamefont {N.}~\bibnamefont {Li}}, \bibinfo
  {author} {\bibfnamefont {S.}~\bibnamefont {Kohler}}, \ and\ \bibinfo {author}
  {\bibfnamefont {P.}~\bibnamefont {H{\"{a}}nggi}},\ }\href {\doibase
  10.1103/PhysRevE.80.061115} {\bibfield  {journal} {\bibinfo  {journal} {Phys.
  Rev. E}\ }\textbf {\bibinfo {volume} {80}},\ \bibinfo {pages} {061115}
  (\bibinfo {year} {2009})}\BibitemShut {NoStop}%
\bibitem [{\citenamefont {Wu}\ \emph {et~al.}(2009)\citenamefont {Wu},
  \citenamefont {Yu},\ and\ \citenamefont {Segal}}]{Wu2009}%
  \BibitemOpen
  \bibfield  {author} {\bibinfo {author} {\bibfnamefont {L.-A.}\ \bibnamefont
  {Wu}}, \bibinfo {author} {\bibfnamefont {C.~X.}\ \bibnamefont {Yu}}, \ and\
  \bibinfo {author} {\bibfnamefont {D.}~\bibnamefont {Segal}},\ }\href
  {\doibase 10.1103/PhysRevE.80.041103} {\bibfield  {journal} {\bibinfo
  {journal} {Phys. Rev. E}\ }\textbf {\bibinfo {volume} {80}},\ \bibinfo
  {pages} {041103} (\bibinfo {year} {2009})}\BibitemShut {NoStop}%
\bibitem [{\citenamefont {Kobayashi}\ \emph {et~al.}(2009)\citenamefont
  {Kobayashi}, \citenamefont {Teraoka},\ and\ \citenamefont
  {Terasaki}}]{Kobayashi2009}%
  \BibitemOpen
  \bibfield  {author} {\bibinfo {author} {\bibfnamefont {W.}~\bibnamefont
  {Kobayashi}}, \bibinfo {author} {\bibfnamefont {Y.}~\bibnamefont {Teraoka}},
  \ and\ \bibinfo {author} {\bibfnamefont {I.}~\bibnamefont {Terasaki}},\
  }\href {\doibase 10.1063/1.3253712} {\bibfield  {journal} {\bibinfo
  {journal} {Appl. Phys. Lett.}\ }\textbf {\bibinfo {volume} {95}},\ \bibinfo
  {pages} {171905} (\bibinfo {year} {2009})}\BibitemShut {NoStop}%
\bibitem [{\citenamefont {Kuo}\ and\ \citenamefont {Chang}(2010)}]{Kuo2010}%
  \BibitemOpen
  \bibfield  {author} {\bibinfo {author} {\bibfnamefont {D.~M.-T.}\
  \bibnamefont {Kuo}}\ and\ \bibinfo {author} {\bibfnamefont {Y.-c.}\
  \bibnamefont {Chang}},\ }\href {\doibase 10.1103/PhysRevB.81.205321}
  {\bibfield  {journal} {\bibinfo  {journal} {Phys. Rev. B}\ }\textbf {\bibinfo
  {volume} {81}},\ \bibinfo {pages} {205321} (\bibinfo {year}
  {2010})}\BibitemShut {NoStop}%
\bibitem [{\citenamefont {Roberts}\ and\ \citenamefont
  {Walker}(2011)}]{Roberts2011}%
  \BibitemOpen
  \bibfield  {author} {\bibinfo {author} {\bibfnamefont {N.~A.}\ \bibnamefont
  {Roberts}}\ and\ \bibinfo {author} {\bibfnamefont {D.~G.}\ \bibnamefont
  {Walker}},\ }\href {\doibase 10.1016/j.ijthermalsci.2010.12.004} {\bibfield
  {journal} {\bibinfo  {journal} {Int. J. Therm. Sci.}\ }\textbf {\bibinfo
  {volume} {50}},\ \bibinfo {pages} {648} (\bibinfo {year} {2011})}\BibitemShut
  {NoStop}%
\bibitem [{\citenamefont {Tian}\ \emph {et~al.}(2012)\citenamefont {Tian},
  \citenamefont {Xie}, \citenamefont {Yang}, \citenamefont {Ren}, \citenamefont
  {Zhang}, \citenamefont {Wang}, \citenamefont {Zhou}, \citenamefont {Peng},
  \citenamefont {Wang},\ and\ \citenamefont {Liu}}]{Tian2012}%
  \BibitemOpen
  \bibfield  {author} {\bibinfo {author} {\bibfnamefont {H.}~\bibnamefont
  {Tian}}, \bibinfo {author} {\bibfnamefont {D.}~\bibnamefont {Xie}}, \bibinfo
  {author} {\bibfnamefont {Y.}~\bibnamefont {Yang}}, \bibinfo {author}
  {\bibfnamefont {T.~L.}\ \bibnamefont {Ren}}, \bibinfo {author} {\bibfnamefont
  {G.}~\bibnamefont {Zhang}}, \bibinfo {author} {\bibfnamefont {Y.~F.}\
  \bibnamefont {Wang}}, \bibinfo {author} {\bibfnamefont {C.~J.}\ \bibnamefont
  {Zhou}}, \bibinfo {author} {\bibfnamefont {P.~G.}\ \bibnamefont {Peng}},
  \bibinfo {author} {\bibfnamefont {L.~G.}\ \bibnamefont {Wang}}, \ and\
  \bibinfo {author} {\bibfnamefont {L.~T.}\ \bibnamefont {Liu}},\ }\href
  {\doibase 10.1038/srep00523} {\bibfield  {journal} {\bibinfo  {journal} {Sci.
  Rep.}\ }\textbf {\bibinfo {volume} {2}},\ \bibinfo {pages} {523} (\bibinfo
  {year} {2012})}\BibitemShut {NoStop}%
\bibitem [{\citenamefont {Ren}\ and\ \citenamefont {Zhu}(2013)}]{Ren2013}%
  \BibitemOpen
  \bibfield  {author} {\bibinfo {author} {\bibfnamefont {J.}~\bibnamefont
  {Ren}}\ and\ \bibinfo {author} {\bibfnamefont {J.-X.}\ \bibnamefont {Zhu}},\
  }\href {\doibase 10.1103/PhysRevB.87.241412} {\bibfield  {journal} {\bibinfo
  {journal} {Phys. Rev. B}\ }\textbf {\bibinfo {volume} {87}},\ \bibinfo
  {pages} {241412(R)} (\bibinfo {year} {2013})}\BibitemShut {NoStop}%
\bibitem [{\citenamefont {Giazotto}\ and\ \citenamefont
  {Bergeret}(2013)}]{Giazotto2013}%
  \BibitemOpen
  \bibfield  {author} {\bibinfo {author} {\bibfnamefont {F.}~\bibnamefont
  {Giazotto}}\ and\ \bibinfo {author} {\bibfnamefont {F.~S.}\ \bibnamefont
  {Bergeret}},\ }\href {\doibase 10.1063/1.4846375} {\bibfield  {journal}
  {\bibinfo  {journal} {Appl. Phys. Lett.}\ }\textbf {\bibinfo {volume}
  {103}},\ \bibinfo {pages} {242602} (\bibinfo {year} {2013})}\BibitemShut
  {NoStop}%
\bibitem [{\citenamefont {Mart{\'{i}}nez-P{\'{e}}rez}\ and\ \citenamefont
  {Giazotto}(2013)}]{Martinez-Perez2013}%
  \BibitemOpen
  \bibfield  {author} {\bibinfo {author} {\bibfnamefont {M.~J.}\ \bibnamefont
  {Mart{\'{i}}nez-P{\'{e}}rez}}\ and\ \bibinfo {author} {\bibfnamefont
  {F.}~\bibnamefont {Giazotto}},\ }\href {\doibase 10.1063/1.4804550}
  {\bibfield  {journal} {\bibinfo  {journal} {Appl. Phys. Lett.}\ }\textbf
  {\bibinfo {volume} {102}},\ \bibinfo {pages} {182602} (\bibinfo {year}
  {2013})}\BibitemShut {NoStop}%
\bibitem [{\citenamefont {Meair}\ and\ \citenamefont
  {Jacquod}(2013)}]{Meair2013}%
  \BibitemOpen
  \bibfield  {author} {\bibinfo {author} {\bibfnamefont {J.}~\bibnamefont
  {Meair}}\ and\ \bibinfo {author} {\bibfnamefont {P.}~\bibnamefont
  {Jacquod}},\ }\href {\doibase 10.1088/0953-8984/25/8/082201} {\bibfield
  {journal} {\bibinfo  {journal} {J. Phys. Cond. Mat.}\ }\textbf {\bibinfo
  {volume} {25}},\ \bibinfo {pages} {082201} (\bibinfo {year}
  {2013})}\BibitemShut {NoStop}%
\bibitem [{\citenamefont {S{\'{a}}nchez}\ and\ \citenamefont
  {L{\'{o}}pez}(2013)}]{Sanchez2013}%
  \BibitemOpen
  \bibfield  {author} {\bibinfo {author} {\bibfnamefont {D.}~\bibnamefont
  {S{\'{a}}nchez}}\ and\ \bibinfo {author} {\bibfnamefont {R.}~\bibnamefont
  {L{\'{o}}pez}},\ }\href {\doibase 10.1103/PhysRevLett.110.026804} {\bibfield
  {journal} {\bibinfo  {journal} {Phys. Rev. Lett.}\ }\textbf {\bibinfo
  {volume} {110}},\ \bibinfo {pages} {026804} (\bibinfo {year}
  {2013})}\BibitemShut {NoStop}%
\bibitem [{\citenamefont {Fornieri}\ \emph {et~al.}(2014)\citenamefont
  {Fornieri}, \citenamefont {Mart{\'{i}}nez-P{\'{e}}rez},\ and\ \citenamefont
  {Giazotto}}]{Fornieri2014}%
  \BibitemOpen
  \bibfield  {author} {\bibinfo {author} {\bibfnamefont {A.}~\bibnamefont
  {Fornieri}}, \bibinfo {author} {\bibfnamefont {M.~J.}\ \bibnamefont
  {Mart{\'{i}}nez-P{\'{e}}rez}}, \ and\ \bibinfo {author} {\bibfnamefont
  {F.}~\bibnamefont {Giazotto}},\ }\href {\doibase 10.1063/1.4875917}
  {\bibfield  {journal} {\bibinfo  {journal} {Appl. Phys. Lett.}\ }\textbf
  {\bibinfo {volume} {104}},\ \bibinfo {pages} {183108} (\bibinfo {year}
  {2014})}\BibitemShut {NoStop}%
\bibitem [{\citenamefont {Jing}\ \emph {et~al.}(2015)\citenamefont {Jing},
  \citenamefont {Segal}, \citenamefont {Li},\ and\ \citenamefont
  {Wu}}]{Jing2015}%
  \BibitemOpen
  \bibfield  {author} {\bibinfo {author} {\bibfnamefont {J.}~\bibnamefont
  {Jing}}, \bibinfo {author} {\bibfnamefont {D.}~\bibnamefont {Segal}},
  \bibinfo {author} {\bibfnamefont {B.}~\bibnamefont {Li}}, \ and\ \bibinfo
  {author} {\bibfnamefont {L.-A.}\ \bibnamefont {Wu}},\ }\href {\doibase
  10.1038/srep15332} {\bibfield  {journal} {\bibinfo  {journal} {Sci. Rep.}\
  }\textbf {\bibinfo {volume} {5}},\ \bibinfo {pages} {15332} (\bibinfo {year}
  {2015})}\BibitemShut {NoStop}%
\bibitem [{\citenamefont {Mart{\'{i}}nez-P{\'{e}}rez}\ \emph
  {et~al.}(2015)\citenamefont {Mart{\'{i}}nez-P{\'{e}}rez}, \citenamefont
  {Fornieri},\ and\ \citenamefont {Giazotto}}]{Martinez-Perez2015}%
  \BibitemOpen
  \bibfield  {author} {\bibinfo {author} {\bibfnamefont {M.~J.}\ \bibnamefont
  {Mart{\'{i}}nez-P{\'{e}}rez}}, \bibinfo {author} {\bibfnamefont
  {A.}~\bibnamefont {Fornieri}}, \ and\ \bibinfo {author} {\bibfnamefont
  {F.}~\bibnamefont {Giazotto}},\ }\href {\doibase 10.1038/nnano.2015.11}
  {\bibfield  {journal} {\bibinfo  {journal} {Nat. Nanotechnol.}\ }\textbf
  {\bibinfo {volume} {10}},\ \bibinfo {pages} {303} (\bibinfo {year}
  {2015})}\BibitemShut {NoStop}%
\bibitem [{\citenamefont {Joulain}\ \emph {et~al.}(2016)\citenamefont
  {Joulain}, \citenamefont {Drevillon}, \citenamefont {Ezzahri},\ and\
  \citenamefont {Ordonez-Miranda}}]{Joulain2016}%
  \BibitemOpen
  \bibfield  {author} {\bibinfo {author} {\bibfnamefont {K.}~\bibnamefont
  {Joulain}}, \bibinfo {author} {\bibfnamefont {J.}~\bibnamefont {Drevillon}},
  \bibinfo {author} {\bibfnamefont {Y.}~\bibnamefont {Ezzahri}}, \ and\
  \bibinfo {author} {\bibfnamefont {J.}~\bibnamefont {Ordonez-Miranda}},\
  }\href {\doibase 10.1103/PhysRevLett.116.200601} {\bibfield  {journal}
  {\bibinfo  {journal} {Phys. Rev. Lett.}\ }\textbf {\bibinfo {volume} {116}},\
  \bibinfo {pages} {200601} (\bibinfo {year} {2016})}\BibitemShut {NoStop}%
\bibitem [{\citenamefont {Ordonez-Miranda}\ \emph {et~al.}(2017)\citenamefont
  {Ordonez-Miranda}, \citenamefont {Ezzahri},\ and\ \citenamefont
  {Joulain}}]{Ordonez-Miranda2017}%
  \BibitemOpen
  \bibfield  {author} {\bibinfo {author} {\bibfnamefont {J.}~\bibnamefont
  {Ordonez-Miranda}}, \bibinfo {author} {\bibfnamefont {Y.}~\bibnamefont
  {Ezzahri}}, \ and\ \bibinfo {author} {\bibfnamefont {K.}~\bibnamefont
  {Joulain}},\ }\href {\doibase 10.1103/PhysRevE.95.022128} {\bibfield
  {journal} {\bibinfo  {journal} {Phys. Rev. E}\ }\textbf {\bibinfo {volume}
  {95}},\ \bibinfo {pages} {022128} (\bibinfo {year} {2017})}\BibitemShut
  {NoStop}%
\bibitem [{\citenamefont {Nakai}\ and\ \citenamefont
  {Nagaosa}(2019)}]{Nakai2019}%
  \BibitemOpen
  \bibfield  {author} {\bibinfo {author} {\bibfnamefont {R.}~\bibnamefont
  {Nakai}}\ and\ \bibinfo {author} {\bibfnamefont {N.}~\bibnamefont
  {Nagaosa}},\ }\href {\doibase 10.1103/PhysRevB.99.115201} {\bibfield
  {journal} {\bibinfo  {journal} {Phys. Rev. B}\ }\textbf {\bibinfo {volume}
  {99}},\ \bibinfo {pages} {115201} (\bibinfo {year} {2019})}\BibitemShut
  {NoStop}%
\bibitem [{\citenamefont {Christen}\ and\ \citenamefont
  {B{\"{u}}ttiker}(1996)}]{Christen1996}%
  \BibitemOpen
  \bibfield  {author} {\bibinfo {author} {\bibfnamefont {T.}~\bibnamefont
  {Christen}}\ and\ \bibinfo {author} {\bibfnamefont {M.}~\bibnamefont
  {B{\"{u}}ttiker}},\ }\href {\doibase 10.1209/epl/i1996-00145-8} {\bibfield
  {journal} {\bibinfo  {journal} {Europhys. Lett.}\ }\textbf {\bibinfo {volume}
  {35}},\ \bibinfo {pages} {523} (\bibinfo {year} {1996})}\BibitemShut
  {NoStop}%
\bibitem [{\citenamefont {Whitney}(2013)}]{Whitney2013}%
  \BibitemOpen
  \bibfield  {author} {\bibinfo {author} {\bibfnamefont {R.~S.}\ \bibnamefont
  {Whitney}},\ }\href {\doibase 10.1103/PhysRevB.87.115404} {\bibfield
  {journal} {\bibinfo  {journal} {Phys. Rev. B}\ }\textbf {\bibinfo {volume}
  {87}},\ \bibinfo {pages} {115404} (\bibinfo {year} {2013})}\BibitemShut
  {NoStop}%
\bibitem [{\citenamefont {Haldane}(1981)}]{Haldane1981}%
  \BibitemOpen
  \bibfield  {author} {\bibinfo {author} {\bibfnamefont {F.~D.}\ \bibnamefont
  {Haldane}},\ }\href {\doibase 10.1088/0022-3719/14/19/010} {\bibfield
  {journal} {\bibinfo  {journal} {J. Phys. C}\ }\textbf
  {\bibinfo {volume} {14}},\ \bibinfo {pages} {2585} (\bibinfo {year}
  {1981})}\BibitemShut {NoStop}%
\bibitem [{\citenamefont {Tomonaga}(1950)}]{Tomonaga1950}%
  \BibitemOpen
  \bibfield  {author} {\bibinfo {author} {\bibfnamefont {S.-i.}\ \bibnamefont
  {Tomonaga}},\ }\href {\doibase 10.1143/ptp/5.4.544} {\bibfield  {journal}
  {\bibinfo  {journal} {Prog. Theor. Phys.}\ }\textbf {\bibinfo {volume} {5}},\
  \bibinfo {pages} {544} (\bibinfo {year} {1950})}\BibitemShut {NoStop}%
\bibitem [{\citenamefont {Luttinger}(1963)}]{Luttinger1963}%
  \BibitemOpen
  \bibfield  {author} {\bibinfo {author} {\bibfnamefont {J.~M.}\ \bibnamefont
  {Luttinger}},\ }\href {\doibase 10.1063/1.1704046} {\bibfield  {journal}
  {\bibinfo  {journal} {J. Math. Phys.}\ }\textbf {\bibinfo {volume} {4}},\
  \bibinfo {pages} {1154} (\bibinfo {year} {1963})}\BibitemShut {NoStop}%
\bibitem [{\citenamefont {Gogolin}\ \emph {et~al.}(1998)\citenamefont
  {Gogolin}, \citenamefont {Nersesyan},\ and\ \citenamefont
  {Tsvelik}}]{Gogolin1998}%
  \BibitemOpen
  \bibfield  {author} {\bibinfo {author} {\bibfnamefont {A.~O.}\ \bibnamefont
  {Gogolin}}, \bibinfo {author} {\bibfnamefont {A.~A.}\ \bibnamefont
  {Nersesyan}}, \ and\ \bibinfo {author} {\bibfnamefont {A.~M.}\ \bibnamefont
  {Tsvelik}},\ }\href@noop {} {\emph {\bibinfo {title} {{Bosonization and
  Strongly Correlated Systems}}}}\ (\bibinfo  {publisher} {Cambridge University
  Press, Cambridge},\ \bibinfo {year} {1998})\BibitemShut {NoStop}%
\bibitem [{\citenamefont {Giamarchi}(2007)}]{Giamarchi2007}%
  \BibitemOpen
  \bibfield  {author} {\bibinfo {author} {\bibfnamefont {T.}~\bibnamefont
  {Giamarchi}},\ }\href {\doibase 10.1093/acprof:oso/9780198525004.001.0001}
  {\emph {\bibinfo {title} {{Quantum Physics in One Dimension}}}}\ (\bibinfo
  {publisher} {Oxford University Press, Oxford},\ \bibinfo {year} {2007})\BibitemShut
  {NoStop}%
\bibitem [{\citenamefont {Egger}\ and\ \citenamefont
  {Grabert}(1998)}]{Egger1998}%
  \BibitemOpen
  \bibfield  {author} {\bibinfo {author} {\bibfnamefont {R.}~\bibnamefont
  {Egger}}\ and\ \bibinfo {author} {\bibfnamefont {H.}~\bibnamefont
  {Grabert}},\ }\href {\doibase 10.1103/PhysRevB.58.10761} {\bibfield
  {journal} {\bibinfo  {journal} {Phys. Rev. B}\ }\textbf {\bibinfo {volume}
  {58}},\ \bibinfo {pages} {10761} (\bibinfo {year} {1998})}\BibitemShut
  {NoStop}%
\bibitem [{\citenamefont {Fisher}\ and\ \citenamefont
  {Glazman}(1997)}]{Fisher1997}%
  \BibitemOpen
  \bibfield  {author} {\bibinfo {author} {\bibfnamefont {M.~P.~A.}\
  \bibnamefont {Fisher}}\ and\ \bibinfo {author} {\bibfnamefont {L.~I.}\
  \bibnamefont {Glazman}},\ }in\ \href {\doibase 10.1007/978-94-015-8839-3_9}
  {\emph {\bibinfo {booktitle} {Mesoscopic Electron Transport}}}\ (\bibinfo
  {publisher} {Springer, Dordrecht},\ \bibinfo {year} {1997})\ pp.\ \bibinfo
  {pages} {331--373}\BibitemShut {NoStop}%
\bibitem [{\citenamefont {Hossein-Nejad}\ \emph {et~al.}(2015)\citenamefont
  {Hossein-Nejad}, \citenamefont {O'Reilly},\ and\ \citenamefont
  {Olaya-Castro}}]{Hossein-Nejad2015}%
  \BibitemOpen
  \bibfield  {author} {\bibinfo {author} {\bibfnamefont {H.}~\bibnamefont
  {Hossein-Nejad}}, \bibinfo {author} {\bibfnamefont {E.~J.}\ \bibnamefont
  {O'Reilly}}, \ and\ \bibinfo {author} {\bibfnamefont {A.}~\bibnamefont
  {Olaya-Castro}},\ }\href {\doibase 10.1088/1367-2630/17/7/075014} {\bibfield
  {journal} {\bibinfo  {journal} {New J. Phys.}\ }\textbf {\bibinfo {volume}
  {17}},\ \bibinfo {pages} {075014} (\bibinfo {year} {2015})}\BibitemShut
  {NoStop}%
\bibitem [{\citenamefont {Weimer}\ \emph {et~al.}(2008)\citenamefont {Weimer},
  \citenamefont {Henrich}, \citenamefont {Rempp}, \citenamefont
  {Schr{\"{o}}der},\ and\ \citenamefont {Mahler}}]{Weimer2008}%
  \BibitemOpen
  \bibfield  {author} {\bibinfo {author} {\bibfnamefont {H.}~\bibnamefont
  {Weimer}}, \bibinfo {author} {\bibfnamefont {M.~J.}\ \bibnamefont {Henrich}},
  \bibinfo {author} {\bibfnamefont {F.}~\bibnamefont {Rempp}}, \bibinfo
  {author} {\bibfnamefont {H.}~\bibnamefont {Schr{\"{o}}der}}, \ and\ \bibinfo
  {author} {\bibfnamefont {G.}~\bibnamefont {Mahler}},\ }\href {\doibase
  10.1209/0295-5075/83/30008} {\bibfield  {journal} {\bibinfo  {journal} {EPL}\
  }\textbf {\bibinfo {volume} {83}},\ \bibinfo {pages} {30008} (\bibinfo {year}
  {2008})}\BibitemShut {NoStop}%
\bibitem [{\citenamefont {Kane}\ and\ \citenamefont {Fisher}(1996)}]{Kane1996}%
  \BibitemOpen
  \bibfield  {author} {\bibinfo {author} {\bibfnamefont {C.~L.}\ \bibnamefont
  {Kane}}\ and\ \bibinfo {author} {\bibfnamefont {M.~P.}\ \bibnamefont
  {Fisher}},\ }\href {\doibase 10.1103/PhysRevLett.76.3192} {\bibfield
  {journal} {\bibinfo  {journal} {Phys. Rev. Lett.}\ }\textbf {\bibinfo
  {volume} {76}},\ \bibinfo {pages} {3192} (\bibinfo {year}
  {1996})}\BibitemShut {NoStop}%
\bibitem [{\citenamefont {Nozi{\`{e}}res}\ and\ \citenamefont {{De
  Dominicis}}(1969)}]{Nozieres1969}%
  \BibitemOpen
  \bibfield  {author} {\bibinfo {author} {\bibfnamefont {P.}~\bibnamefont
  {Nozi{\`{e}}res}}\ and\ \bibinfo {author} {\bibfnamefont {C.~T.}\
  \bibnamefont {{De Dominicis}}},\ }\href {\doibase 10.1103/PhysRev.178.1097}
  {\bibfield  {journal} {\bibinfo  {journal} {Phys. Rev.}\ }\textbf {\bibinfo
  {volume} {178}},\ \bibinfo {pages} {1097} (\bibinfo {year}
  {1969})}\BibitemShut {NoStop}%
\bibitem [{\citenamefont {Auslaender}\ \emph {et~al.}(2002)\citenamefont
  {Auslaender}, \citenamefont {Yacoby}, \citenamefont {{De Picciotto}},
  \citenamefont {Baldwin}, \citenamefont {Pfeiffer},\ and\ \citenamefont
  {West}}]{Auslaender2002}%
  \BibitemOpen
  \bibfield  {author} {\bibinfo {author} {\bibfnamefont {O.~M.}\ \bibnamefont
  {Auslaender}}, \bibinfo {author} {\bibfnamefont {A.}~\bibnamefont {Yacoby}},
  \bibinfo {author} {\bibfnamefont {R.}~\bibnamefont {{De Picciotto}}},
  \bibinfo {author} {\bibfnamefont {K.~W.}\ \bibnamefont {Baldwin}}, \bibinfo
  {author} {\bibfnamefont {L.~N.}\ \bibnamefont {Pfeiffer}}, \ and\ \bibinfo
  {author} {\bibfnamefont {K.~W.}\ \bibnamefont {West}},\ }\href {\doibase
  10.1126/science.1066266} {\bibfield  {journal} {\bibinfo  {journal}
  {Science}\ }\textbf {\bibinfo {volume} {295}},\ \bibinfo {pages} {825}
  (\bibinfo {year} {2002})}\BibitemShut {NoStop}%
\bibitem [{\citenamefont {Steinberg}\ \emph {et~al.}(2008)\citenamefont
  {Steinberg}, \citenamefont {Barak}, \citenamefont {Yacoby}, \citenamefont
  {Pfeiffer}, \citenamefont {West}, \citenamefont {Halperin},\ and\
  \citenamefont {{Le Hur}}}]{Steinberg2008}%
  \BibitemOpen
  \bibfield  {author} {\bibinfo {author} {\bibfnamefont {H.}~\bibnamefont
  {Steinberg}}, \bibinfo {author} {\bibfnamefont {G.}~\bibnamefont {Barak}},
  \bibinfo {author} {\bibfnamefont {A.}~\bibnamefont {Yacoby}}, \bibinfo
  {author} {\bibfnamefont {L.~N.}\ \bibnamefont {Pfeiffer}}, \bibinfo {author}
  {\bibfnamefont {K.~W.}\ \bibnamefont {West}}, \bibinfo {author}
  {\bibfnamefont {B.~I.}\ \bibnamefont {Halperin}}, \ and\ \bibinfo {author}
  {\bibfnamefont {K.}~\bibnamefont {{Le Hur}}},\ }\href {\doibase
  10.1038/nphys810} {\bibfield  {journal} {\bibinfo  {journal} {Nat. Phys.}\
  }\textbf {\bibinfo {volume} {4}},\ \bibinfo {pages} {116} (\bibinfo {year}
  {2008})}\BibitemShut {NoStop}%
\bibitem [{\citenamefont {Scheller}\ \emph {et~al.}(2014)\citenamefont
  {Scheller}, \citenamefont {Liu}, \citenamefont {Barak}, \citenamefont
  {Yacoby}, \citenamefont {Pfeiffer}, \citenamefont {West},\ and\ \citenamefont
  {Zumb{\"{u}}hl}}]{Scheller2014}%
  \BibitemOpen
  \bibfield  {author} {\bibinfo {author} {\bibfnamefont {C.~P.}\ \bibnamefont
  {Scheller}}, \bibinfo {author} {\bibfnamefont {T.~M.}\ \bibnamefont {Liu}},
  \bibinfo {author} {\bibfnamefont {G.}~\bibnamefont {Barak}}, \bibinfo
  {author} {\bibfnamefont {A.}~\bibnamefont {Yacoby}}, \bibinfo {author}
  {\bibfnamefont {L.~N.}\ \bibnamefont {Pfeiffer}}, \bibinfo {author}
  {\bibfnamefont {K.~W.}\ \bibnamefont {West}}, \ and\ \bibinfo {author}
  {\bibfnamefont {D.~M.}\ \bibnamefont {Zumb{\"{u}}hl}},\ }\href {\doibase
  10.1103/PhysRevLett.112.066801} {\bibfield  {journal} {\bibinfo  {journal}
  {Phys. Rev. Lett.}\ }\textbf {\bibinfo {volume} {112}},\ \bibinfo {pages}
  {066801} (\bibinfo {year} {2014})}\BibitemShut {NoStop}%
\end{thebibliography}
%


\end{document}